\def\rn{\noindent\parshape 2 0truecm 8.5truecm 0.3truecm 8.2truecm}
\def\rn{}
\def\nn#1 #2{#1, #2.}				
\def\nnn#1 #2 #3{#1, #2. #3.}			
\def\nnnn#1 #2 #3 #4{#1, #2. #3. #4.}		
\def\nnnnn#1 #2 #3 #4 #5{#1, #2. #3. #4. #5.}	
\def\dualand{ and\hbox{ }}				
\def\multiand{, and\hbox{ }}				
\def\rf#1;#2;#3;#4;#5 {\par\rn#1 #2, {\it #3}, {\bf #4}, #5\par}
\def\rg#1;#2;#3;#4;#5;#6 {\par\rn#1 #2, {\it #3}, {\bf #4}, #5 (``#6'') \par}
\def\rfbook#1;#2;#3;#4;#5 {{\frenchspacing\par\rn#1 #2, {\it #3} (#4: #5)\par}}
\def\rfproc#1;#2;#3;#4;#5;#6 {{\frenchspacing\par\rn#1 #2, in {\it #3}, ed. #4 (#5: #6)\par}}
\def\rfprocp#1;#2;#3;#4;#5;#6;#7 {{\frenchspacing\par\rn#1 #2, in {\it #3}, ed. #4 (#5: #6), p#7\par}}
\def\rfprep#1;#2;#3  {{\par\rn#1 #2, #3\par}}

\def\K{{\rm K}}

\def\Mpc{{\rm Mpc}}
\def\MHz{{\rm MHz}}

\def\expec#1{\langle#1\rangle}

\def\etal{{\frenchspacing\it et al.}}

\def\eg{{\frenchspacing\it e.g.}}
\def\etc{{\frenchspacing\it etc.}}

\def\beq#1{\begin{equation}\label{#1}}
\def\eeq{\end{equation}}
\def\beqa#1{\begin{eqnarray}\label{#1}}
\def\eeqa{\end{eqnarray}}
\def\eq#1{equation~(\ref{#1})}

\def\fig#1{Figure~\ref{#1}}
\def\Fig#1{Figure~\ref{#1}}

\def\sec#1{Section~\ref{#1}}

\def\subsec#1{Subsection~\ref{#1}}

\def\spose#1{\hbox to 0pt{#1\hss}}
\def\simlt{\mathrel{\spose{\lower 3pt\hbox{$\mathchar"218$}}
     \raise 2.0pt\hbox{$\mathchar"13C$}}}
\def\simgt{\mathrel{\spose{\lower 3pt\hbox{$\mathchar"218$}}
     \raise 2.0pt\hbox{$\mathchar"13E$}}}
\def\simpropto{\mathrel{\spose{\lower 3pt\hbox{$\mathchar"218$}}
     \raise 2.0pt\hbox{$\propto$}}}

\def\ed{\end{document}}

\def\Ob{\Omega_{\rm b}}

\def\Om{\Omega_{\rm m}}

\def\rhom{\rho_{{\rm  m}0}}

\def\xe{x_e}

\def\dt{\Delta\theta}
\def\dn{\Delta\nu}

\def\pd{\bot}

\def\sv{\sigma_p}
\def\sb{\sigma_b}

\def\kHz{{\rm kHz}}
\def\km{{\rm km}}

\documentclass{emulateapj}

\begin{document}



\title{Redshift space 21 cm power spectra from reionization}

\author{Xiaomin Wang$^{1}$ and Wayne Hu$^{1,2}$}
\affil{${}^1$Kavli Institute for Cosmological Physics, Univ. of Chicago, Chicago, IL 60637 \\
${}^2$ Department of Astronomy and Astrophysics, Enrico Fermi Institute, Univ. of Chicago, 
Chicago, IL 60637\\
xiaomin@cfcp.uchicago.edu, whu@background.uchicago.edu}


\begin{abstract} 

We construct a simple but self-consistent analytic ionization model for rapid exploration 
of  21cm power spectrum observables in redshift space. It is fully
described by the average ionization fraction $x_e(z)$ and HII patch size $R(z)$ 
and has the flexibility to
accommodate various reionization scenarios. The model associates
ionization regions with dark matter halos of the number density required to recover $x_e$ and
treats redshift space distortions self-consistently with the virial velocity of such halos.
 Based on this model, 
we study the line-of-sight structures in the brightness fluctuations since they are the
most immune to foreground contamination.   
We explore the degeneracy between the HII patch size and nonlinear redshift space distortion in
the one dimensional power spectrum.  We also discuss the limitations experimental
frequency and angular resolutions place on their distinguishability.  
Angular resolution dilutes even the radial signal and will be a serious limitation for
resolving small bubbles before the end of reionization.  Nonlinear redshift space
distortions suggest that a resolution of order 1 -- 10\arcsec\ and a frequency resolution of 
10kHz will ultimately be desirable to extract the full information in the radial field at $z\sim 10$.
First generation instruments
such as LOFAR and MWA can potentially 
measure radial HII patches of a few comoving Mpc 
and larger at the end of reionization and are unlikely to be affected by nonlinear
redshift space distortions.
\end{abstract}



\section{Introduction}

The epoch of reionization is perhaps one of the ``darkest" place in modern cosmology nowadays. 
On one hand, Gunn-Peterson constraints \citep{GunnPeterson} from quasars \citep[\eg,][]{Becker01,Fan02}
 tells us that the Universe was mostly ionized by $z\sim 6$. 
On the other hand, the optical depth to Thomson scattering
 $\tau\sim 0.17$ from WMAP data \citep{Kogut03,Spergel03} 
implies that  reionization may have begun as early as redshift $z\sim 17$. 
While these data suggest reionization was probably an extended and complicated
process,   we still have little information about what exactly happened during this period.   

Among various approaches to explore the reionization epoch, 21cm radiation from neutral hydrogen
is in principle on of the most powerful tools \citep[e.g.][]{HogRee79,ScoRee90,MadMeiRee97}
and  has recently received much attention  
 in preparation for
a new generation of experiments  \citep[\eg,][] 
{BarkanaLoeb04a,Carilli05,ChenME04,CiardiMadau03,Furlanetto03,FZH04,Furlanetto05,
Morales04,MoralesHewitt03,Pen04,Santos04,WyitheLoeb04,Matias03}.
Although contamination from foregrounds will likely dominate the total signal, their smoothness in frequency should allow the measurement of radial structures in the brightness fluctuations.

 Radial structures however are measured in redshift space
and are distorted by the peculiar velocity of
the neutral gas.  
Recent studies have addressed the impact of linear velocity flows on 21cm power spectra
 \citep{Barkana05,BarkanaLoeb04a,BarkanaLoeb04b,BharadwajAli05,Cooray04,DesNusser04}.

In this paper, we provide a complete treatment including the non-linear redshift space
distortions or ``Fingers of God" that blur out radial structures along the line of sight.We show how these effects impact 21cm
power spectra, how they manifest themselves with different experimental setups, 
and how they interact with intrinsic
reionization properties such as the size of the HII regions. 
We show that nonlinear redshift space distortions will ultimately limit our ability
to study small HII structures during reionization.

The outline of the paper is as follows.  
In \sec{prsec}, we construct a simple but self-consistent analytic 
ionization model based on an association with dark matter halos.  Our model is
similar in construction 
to that of \cite{FZH04a} but allows for a rapid exploration of the possible parameter
space of alternate models.    We discuss a generalization of our model that includes
a HII region size distribution in the Appendix.
In \sec{pssec}, we employ the association with dark matter halos to study redshift
space distortions  based on their bias
and virial velocity. 
In \sec{psdtdnsec}, we study the impact of experimental 
angular and frequency resolution on resolving the features induced by the HII
regions and redshift space distortions. 
In \sec{expsec}, we discuss the impact of these considerations on 
first generation experiments such as MWA\footnote{http://web.haystack.mit.edu/MWA/MWA.html}  and LOFAR\footnote{http://www.lofar.org} as well as the design of an experiment that
can recover essentially all of the information in the radial temperature field. 
We summarize our results in \sec{discussionsec}. 

\vfill

\section{Real space 21cm power spectrum}
\label{prsec}
\subsection{Brightness temperature}

The redshifted brightness temperature fluctuation in the 21cm line
can be viewed as a three dimensional field where the position is
specified by the angular coordinates $\vec{\theta}$ on a small
patch of the sky and the observation frequency
$\nu$.  In the absence of redshift space distortions,
 the mapping to comoving coordinates
\begin{equation}
{\vec r} = D_\perp (z) \vec{\theta} +   D(z) \vec{e}_{r}
\end{equation}
involves the redshift $z= \nu_{21}/\nu -1$, where $\nu_{21}=1420.4$MHz,
the comoving transverse (or angular diameter) distance  $D_{\perp}(z)$,
and the comoving distance along the radial direction $\vec{e}_{r}$, $D(z)$.
In a flat universe, which we assume throughout,
$D_\perp=D$.

The amplitude of the brightness temperature fluctuation is proportional to 
the neutral hydrogen density fluctuation $\delta_{\rm HI} \equiv \delta n_{\rm HI}/
\bar n_{\rm H}$
\begin{equation}
\delta T({\vec r}) = F^{1/2}(z) \delta_{\rm HI}({\vec r})\,,
\end{equation}
where $\bar n_{\rm H}$ is the mean hydrogen density.
Under the assumption that the spin temperature is much larger than the CMB temperature,
the proportionality coefficient \citep{HogRee79}
\begin{eqnarray}
\label{factoreq}
F^{1/2}(z)&=& {3 \over 16} {c^3 \hbar A_{21} \bar n_{\rm H} \over k \nu_{21}^2 H(z)}{1\over 1+z}
\nonumber \\
&\approx&
0.023\K {{1-Y_p \over 0.75}} 
 {{\Ob h^2}\over 0.02} \left( {{1+z}\over 10}{0.15\over \Om h^2} \right)^{1/2} \,.
\end{eqnarray}
Here $A_{21}$  is the Einstein coefficient for spontaneous emission and $Y_p$ is the
primordial helium mass fraction.

Although the mapping involves integrals over the expansion rate $H(z)$, which is currently uncertain
at low redshifts due to the dark energy, at high redshifts it is already well determined by CMB observations.
We will consider a small range of observing frequencies around a central value $z \gg 1$.
Differences in observing frequency are then simply proportional 
to differences in radial distance
\begin{eqnarray}
{\delta D} = {\delta z \over H(z)} \approx 17 {\rm Mpc} \left( {1+z\over 10} {0.15 \over \Om h^2} 
\right)^{1/2} {\delta \nu \over 1 {\rm MHz}} \,.
\label{eqn:radialdistance}
\end{eqnarray}
We set $c=1$ where no confusion will arise.
The transverse distance in a flat universe can be scaled forward from the CMB determined
distance to recombination $D_* \equiv D_\perp(z_*)$ to avoid ambiguities due to the
dark energy
\begin{eqnarray}
D_\perp & = &  D_* -  \int_{z}^{z_*} {d z \over H(z)} \nonumber\\
&\approx & D_* -
4.90 {\rm Gpc} \left( {0.15 \over \Om h^2}{10 \over 1+z}\right)^{1/2} \nonumber\\
&& \times \left[ 1 - \left( { 1+z \over 1+z_*} + {1+z \over 1 + z_{\rm eq}} \right)^{1/2}  \right]\,.
\label{eqn:angulardiameter}
\end{eqnarray}
where the redshift of matter radiation equality is given by
\begin{eqnarray}
1+z_{\rm eq} = 3600 \left( { \Om h^2 \over 0.15 }\right) \,.
 \end{eqnarray}
Notice there is no dependence on and no sensitivity to the dark energy.

 Likewise the conversion factor $F$
 from neutral hydrogen density to brightness fluctuation 
 only adds a sensitivity to the baryon density $\Ob h^2$.  
  Constraints on these quantities from the first year WMAP data \citep{Spergel03}, require
$\Om h^2=0.14\pm 0.2$ and
$\Ob h^2= 0.024\pm 0.001$  which 
 corresponds to
  $D_* = 13.7 \pm 0.5$ Gpc, $z_* = 1088^{+1}_{-2}$.
For definiteness and in accordance with 
large scale structure data sets \citep{sdsslyaf,sdsspars}, we will illustrate our results in a fiducial  
  $\Lambda$CDM
 cosmology with $\Om = 1-\Omega_\Lambda =0.29$, $\Omega_b h^2 =0.0244$,
 $\Om  h^2=0.15$.  We assume that the initial spectrum of curvature perturbations $\zeta$ takes
  the form
 \begin{equation}
 {k^3 \over 2\pi^2} P_\zeta  = \delta_{\zeta}^2 \left({k \over 0.05 {\rm Mpc}^{-1}} \right)^{n-1} \
 \end{equation}
 with
 $\delta_\zeta=4.8 \times 10^{-5}$ and $n=0.99$, corresponding to $\sigma_8 = 0.9$.
 This requires $D_* = 13.4 {\rm Gpc}$, $z_*=1088$ and 
  the total Thomson optical depth during reionization to be $\tau=0.103$ or
 a lower limit on the redshift of the beginning of reionization of $z=12$.  
 In the rest of this section, we shall treat the 
 observations as if they were made directly in 3D physical space on the neutral hydrogen
 density field. 
 
\smallskip   
  
\subsection{Neutral hydrogen halo model}

The neutral hydrogen fluctuations can be separated into contributions from
the ionization fraction $x({\vec r})= x_e + \delta x({\vec r})$ and gas density fluctuations $\delta(\vec r)$
\begin{equation}
\delta_{\rm HI}(\vec r) = [1-x_e - \delta x(\vec r)] \delta(\vec r) - \delta x({\vec r}) \,.
\end{equation}
The two point correlation function of the neutral hydrogen field then becomes
\citep{FZH04}
\begin{eqnarray}
\xi_{\delta_{\rm HI} \delta_{\rm HI}}(r) 
&=& \xi_{\delta x \delta x}(r) - 2(1-x_e) \xi_{\delta x \delta}(r) \nonumber\\
&& + (1-x_e)^2 \xi_{\delta \delta}(r)
+ \xi_{\delta x \delta \, \delta x \delta}(r) \,.
\label{eqn:correlationfunction}
\end{eqnarray}
Here we have employed the notation
\begin{equation}
\xi_{ab}(| \vec{r}_1 - \vec{r}_2| ) = \langle a(\vec{r}_1)  b(\vec{r}_2) \rangle
\end{equation}
 for the two point correlation function between field $a$ and $b$.  Analogously
 we define the power spectra as the Fourier transform of the correlation functions
 \begin{equation}
 P_{ab}(k) = \int {d^3 k \over (2\pi)^3} e^{i \vec{k} \cdot \vec{r}} \xi_{ab}(r) \,.
 \end{equation}
 It is also useful to define the 1D line of sight 
 projected power spectrum since the radial structure of the field is critical for removing
 foreground contaminants.  In general it is defined as 
\begin{eqnarray}
P^{\rm 1D}_{ab}(k_\parallel) &=&
\int   { d^2 k_\perp \over (2\pi)^2  }P_{ab}({k}) \nonumber\\
&=& \int_{k_{\parallel}}^{\infty} {k dk \over 2\pi } P_{ab}({k}) \,,
\label{eqn:1Ddef}
\end{eqnarray}
 where $k_\parallel$ and $k_\perp$ are the components of $\vec{k}$ parallel and perpendicular
 to the line of sight.  
  Finally we will occasionally use the shorthand notation $P_{a} \equiv P_{aa}$
 where no confusion will arise.
 
The final term in \eq{eqn:correlationfunction} involves a product of density and
ionization fluctuations.   Though formally second order for linear fluctuations,
we shall see that for density perturbation on scales smaller than a typical
ionized region, this term must
be included to construct a physical model of the neutral hydrogen field because of
contributions of the form  \citep{FZH04}
\begin{equation}
\xi_{\delta x \delta \, \delta x \delta}(r) \approx
\xi_{\delta x \delta x}(r)\xi_{\delta \delta}(r) \,.
\end{equation}

We model the underlying density and ionization correlation functions with a halo model of
ionization bubbles.
Our implementation represents a simplification of the \cite{FZH04} model in that we take a single
characteristic bubble size at each redshift.   On the other hand, we leave the evolution of
the bubble size and mean ionization arbitrary to ensure flexibility.
  For example, in our model  reionization can end by the percolation of many small
bubbles or the growth of rare bubbles until overlap. 
More importantly our model captures, in a simple and fully analytic
form, the required physical scaling of the correlations
as a function of  mean ionization and bubble size \citep[cf.\ ][]{Santos03}.

\begin{figure}[th]
\centerline{\epsfxsize=8.5cm\epsffile{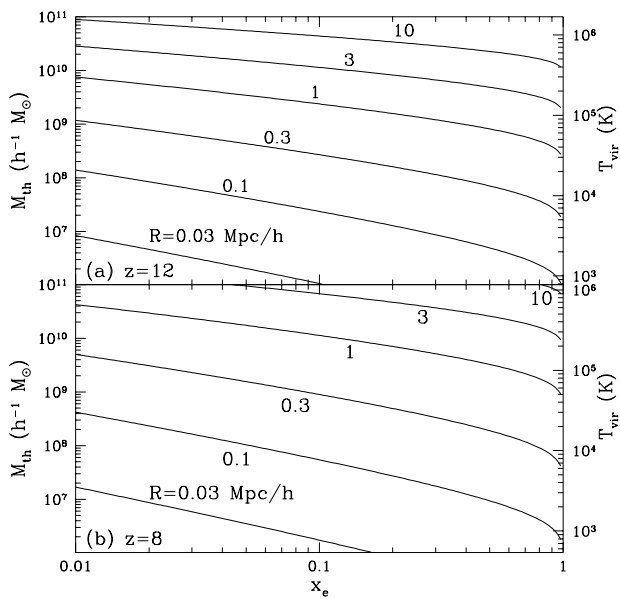}}
\caption{\label{mthxe}\footnotesize
Mass threshold and virial temperature (see \eq{eqn:virial})
 as a function of the mean ionization $x_e$
for several choices of the bubble radius $R$ at $z=12$ and $z=8$.
}
\smallskip
\end{figure}

In our model, the correlation functions are described by two quantities: the typical
radius of ionization bubbles $R(z)$  and the mean ionization fraction $x_e(z)$.  
We assume that the probability that a given point in space is ionized is determined by
a Poisson process such that 
\begin{equation}
\langle x(\vec{r}) \rangle  = 1 - e^{-n_b(\vec{r}) V_b} \,,
\label{eqn:poissonmodel}
\end{equation}
where $n_b$ is the number density of bubbles and $V_b = 4\pi R^3/3$ is the volume of the
bubbles.   Brackets here denote averaging over the Poisson process.
The mean number density is related to the mean ionization by
\begin{equation}
\bar n_b = -{1 \over V_b} \ln (1-x_e) \,.
\end{equation} 
By associating the seeds of the ionization bubbles with massive dark matter
halos of that number density, we can model the statistics of the bubbles.
Specifically we set a mass threshold such that
\begin{equation}
\bar n_b = \int_{M_{\rm th}}^\infty {dM \over M} {d n_h \over d\ln M} \,,
\end{equation}
where $dn_h / d\ln M$ is the halo mass function \citep{SheTor99}
\begin{eqnarray}
{d n_h \over d\ln M}= { \rhom \over M} f(\nu) {d\nu \over d\ln M}\,.
\label{eqn:massfn}
\end{eqnarray}
Here $\rhom$ is the matter density today,
$\nu = \delta_c/\sigma_{\rm lin}(M;z)$ and
\begin{eqnarray}
\nu f(\nu) = A\sqrt{{2 \over \pi} a\nu^2 } [1+(a\nu^2)^{-p}] \exp[-a\nu^2/2]\,,
\label{eqn:stform}
\end{eqnarray} 
and $\sigma_{\rm lin}$ is the rms of the linear density field smoothed by a
tophat of a radius that encloses a mass $M$.
A choice of 
$\delta_c=1.69$, $a=0.75$, $p=0.3$, and $A$ such that $\int d\nu f(\nu)=1$
fits the results of simulations well.
Note that our model can accommodate collective effects where smaller
halos associated with the seed halo can make the bubble radius 
much larger than the virial radius of the seed halo \citep{BarLoe04c}.
In \fig{mthxe},  
we show the mass threshold $M_{\rm th}$ as a function of $x_{e}$
for several choices of the bubble size $R$ and redshift.

\begin{figure}[th]
\centerline{\epsfxsize=8.5cm\epsffile{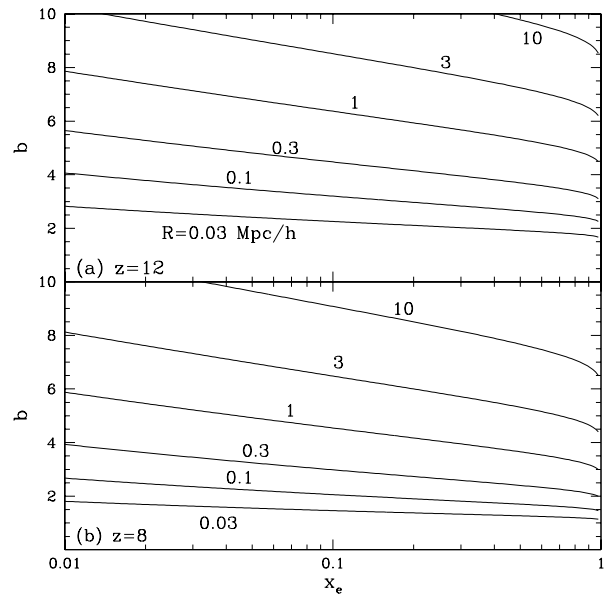}}
\caption{\label{bxe}\footnotesize%
Bubble bias as a function of the mean ionization $x_e$
for several choices of the bubble radius $R$ at $z=12$ and $z=8$.  The bubble
bias increases with $R$ at fixed $x_{e}$ since the bubbles become rare and
highly correlated.}
\end{figure}

\smallskip

\subsection{Two bubble correlations}
\label{sec:twobubble}

For scales where $r \gg R$, the two point functions are dominated
by the correlations between two separate bubbles as in the two halo regime of
the halo model.  These correlations are induced by 
the enhanced probability of bubble formation in over dense regions through 
\eq{eqn:poissonmodel}. 

Fluctuations in the bubble number density will then follow the density fluctuations of 
the dark matter  through
the bubble bias
\begin{equation}
b =  {1 \over \bar n_b} \int_{M_{\rm th}}^\infty {dM \over M} b_h(M) {d n_h \over d\ln M}\,,
\end{equation}
where the halo bias is given by \citep{SheTor99}
\begin{equation}
b_h = 1 + {a \nu^2 -1 \over \delta_c}
         + { 2 p \over \delta_c [ 1 + (a \nu^2)^p]}\,.
\label{eqn:halobias}
\end{equation}
Specifically, we model
\begin{equation}
n_b(\vec{r}) = \bar n_b (1+ b \delta_W) \,,
\end{equation}
where $\delta_{W}$ is the density fluctuation field smoothed by a top hat 
window of radius $R$,
\begin{equation}
\delta_W(\vec{r}) \equiv \int d^3 r'  \delta(\vec{r}')W_{R}(\vec{r}-\vec{r} ')\,,
\end{equation}
where $W_{R}(r) = V_b^{-1}$ for $r\le R$ and 0 otherwise.  
In \fig{bxe}, 
we show the bubble bias as a function of $x_{e}$ for several values of $R$ and $z$.

\begin{figure}[th]
\centerline{\epsfxsize=8.5cm\epsffile{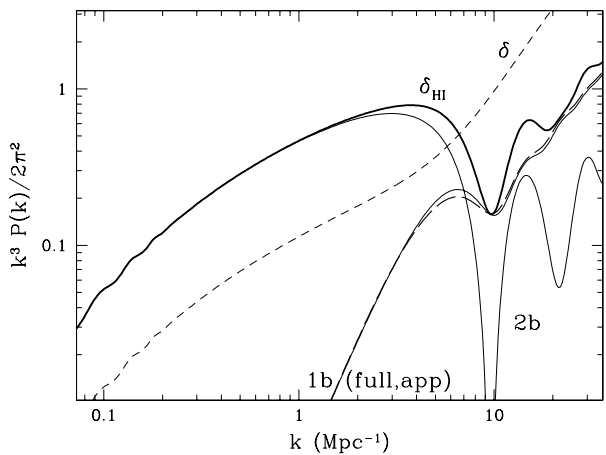}}
\caption{\label{pk}\footnotesize%
Neutral hydrogen power spectrum at $z=8$ with $x_{e}=0.8$ and $R=0.3$Mpc$/h$ broken up into
components from the two bubble and one bubble contributions.  The one bubble contributions
are further shown with the full convolution factor (solid) and the approximation of 
\eq{eqn:convolutionapprox} (long dashed).  Oscillations at high $k$ are an artifact of taking a single bubble
scale with a top hat profile and should be removed by averaging (see Appendix) or projection. 
For reference the density power spectrum is also
shown (short dashed). }
\end{figure}

Expanding \eq{eqn:poissonmodel} 
in the density fluctuations and taking the Fourier transforms yields
\begin{eqnarray}
P_{\delta x \delta x}^{2b}(k)& =& [(1-x_e) \ln (1-x_e) b W_{R}(k) ]^2 P_{\delta}(k) \,,\nonumber\\
P_{\delta x \delta}^{2b}(k) & = & - (1-x_e) \ln (1-x_e) b W_{R}(k) P_{\delta}(k)\,,
\label{eqn:twobubble}
\end{eqnarray}
where $W_{R}(k)$ is the Fourier transform of the top hat window and the superscript $2b$
denotes two bubble contributions.  Here we have assumed that the gas density fluctuations
trace the dark matter fluctuations in the two bubble regime.
Furthermore, in this regime $P_{\delta x \delta \, \delta x \delta}^{2b}$ is second order in
the density fluctuation and we therefore set it to zero. 
The power spectrum of the two bubble contributions then becomes
\begin{eqnarray}
P_{\delta_{\rm HI}}^{2b}(k) = (1- x_{e})^{2} b_{\rm eff}^{2} P_{\delta}(k)\,,
\label{eqn:p2b}
\end{eqnarray}
where the total effective bias includes both density and ionization fluctuations
\begin{equation}
 b_{\rm eff} = \ln(1-x_{e})b W_{R} + 1\,.
\end{equation}
Note that this expression has the proper limiting behavior. As the mean ionization
$x_e\rightarrow 0$, the ionization
bubbles disappear, $b_{\rm eff} \rightarrow 1$ and $P_{\delta_{\rm HI}}^{2b} \rightarrow P_{\delta}$.
As $x_{e }\rightarrow 1$ all of the neutral hydrogen disappears and $P_{\delta_{\rm HI}}^{2b} 
\rightarrow 0$.  Note that this behavior is independent of $R$.  Full reionization can occur
either by the growth of small bubbles into large ones or the creation of a high number density
of overlapping small bubbles.  Finally the effective bias $b_{\rm eff}$
of the neutral hydrogen field is typically negative in the two bubble regime given a model
such as ours where ionization occurs in over dense regions.
  An example of the two bubble power spectrum is shown
in \fig{pk}.

\smallskip

\subsection{One bubble correlations}

For scales where $r \ll R$, the correlation functions are dominated by the presence
or absence of a single bubble.   Two points separated
by $r \ll R$ are either a part of an ionization bubble or not and hence the ionization 
probability 
must converge to $x_e$.  Two points separated by $r \gg R$ have their ionization probability
converge to $x_e^2$.  Hence the full correlation function must interpolate between the
two \citep{GruHu98}
\begin{eqnarray}
\langle x(\vec{r}_1)  x(\vec{r}_2) \rangle = x_e^2 + (x_e - x_e^2) f(r/R)\,.
\label{eqn:corrform}
\end{eqnarray}
The interpolation function $f(r/R)$ must satisfy
\begin{eqnarray}
\lim_{r \ll R} f(r/R) &\rightarrow& 1 \,, \nonumber\\
\lim_{r \gg R} f(r/R) & \rightarrow& 0 \,.
\end{eqnarray}
Therefore the ionization fluctuations become
\begin{equation}
\xi_{\delta x\delta x}^{1b} = (x_e - x_e^2) f(r/R) \,.
\end{equation}
Since inside a bubble the medium is fully ionized, the density ionization cross correlation
is negligible in this regime $\xi_{\delta x \delta}^{1b} \approx 0$.  Note however that
$\xi_{\delta x\delta  \, \delta x \delta}^{1b}$ is of the same order as
$\xi_{\delta\delta}$ and can not
be neglected.

To determine $f(r/R)$ for our assumed top-hat ionization bubbles, we consider the
case where bubble overlap is negligible. 
By analogy to the one halo regime of the familiar halo model  \citep[see e.g.][]{CooShe01}
where the halo density profile is replaced with the ionization profile, we can
immediately obtain 
$f(r/R)$ as the convolution of two top-hat windows $V_b W_R(r)$
\begin{equation}
\begin{array}{lll}
f(r/R) &= 1 -\frac{3}{4} \left( {r \over R}\right) + {1\over 16} \left( {r \over R} \right)^3 \,, & r \le 2R \,,
\nonumber \\
  &= 0\,, & r > 2R \,.
 \end{array}
\end{equation}
We have verified through the procedure outlined in  \cite{FZH04} that this remains a reasonable
approximation for the overlap regime as it simply represents an interpolation between the two
physical limits imposed by 
\eq{eqn:corrform} 
through the scale $R$.

Summing the contributions in 
\eq{eqn:correlationfunction} 
and taking the
the Fourier transform, we obtain the one bubble contributions to the power spectrum
\begin{equation}
P_{\delta_{\rm HI}}^{1b}(k) =  ( x_{e} -  x_{e}^{2}) \left[  V_{b}W^{2}_R(k) + \tilde P_{\delta}(k) \right]\,,
\label{eqn:onebubble}
\end{equation}
where 
\begin{equation}
\tilde P_{\delta}(k) = V_{b} \int {d^{3}k_{1} \over (2\pi)^{3}} W^{2}_R(k_{1}) P_{\delta}(\vec{k} - \vec{k}_{1}) \,.
\end{equation}
This construction guarantees the proper limiting behavior
of $P_{\delta_{\rm HI}}^{1b} \rightarrow 0$
as $x_e \rightarrow 0$ and $x_e\rightarrow 1$.
Note that we have arbitrarily associated the density fluctuation term with the two bubble contributions
so that $P_{\delta}^{1b}=0$.  

The first term in 
\eq{eqn:onebubble} 
represents the shot noise contributions 
of the bubbles.   In the limit where bubbles
overlap negligibly $x_e \approx \bar n_b V_b \gg x_e^2$, the first term becomes $x_e^2/ \bar n_b$
as expected.  The second term comes from $\xi_{\delta x\delta x}(r) \xi_{\delta\delta}(r)$ and
involves a convolution of power spectra.  Note that
\begin{equation}
\lim_{kR \gg 1} \tilde P_{\delta}(k) \approx P_{\delta}(k) V_{b} \int {d^{3}k_{1} \over (2\pi)^{3}}  W^{2}_R(k_{1}) = P_{\delta}(k) \,.
\end{equation}
On small scales the total contribution involving the density fluctuations directly become
\begin{equation}
(1-x_e)^2 P_{\delta}(k) +  (x_e - x_e^2)  P^{}_{\delta}(k) = (1- x_{e})P_{\delta}(k) \,.
\end{equation}
Again, the reason is that for $r \ll R$ the region is either ionized or neutral so that density fluctuations contribute
to neutral hydrogen fluctuations with only one factor of 
 $(1- x_{e})$ representing the volume fraction that is
neutral.   Without this term, the neutral hydrogen power spectrum would be unphysical 
in the one bubble regime. 
In the opposite limit,
\begin{eqnarray}
\lim_{kR \ll 1} \tilde P_{\delta}(k) &\approx& V_{b} \int {d^{3}k_1 \over (2\pi)^{3}}
W^{2}_R(k_1) P_{\delta}(k_1) \nonumber\\ 
&= &V_b \sigma_R^2 \,,
\label{eqn:convolvelimit}
\end{eqnarray}
where $\sigma_R^2$ is the variance of the density field smoothed by a tophat
on a scale $R$.   We find that
a simple interpolation between the two regimes  \begin{equation}
 \tilde P_{\delta}(k) = {P_{\delta}(k) V_b \sigma_R^2 \over [ (P_{\delta}(k))^2 + (V_b \sigma_R^2)^2]^{1/2}} \,,
 \label{eqn:convolutionapprox}
 \end{equation}
suffices for obtaining power spectra to 10-20\%
accuracy 
(see \fig{pk}).

\smallskip

\subsection{Fiducial model}
\label{sec:fiducial}

The total neutral hydrogen power spectrum is simply the sum of the one bubble and two bubble terms
in \eq{eqn:p2b} and \eq{eqn:onebubble}.  
The 21 cm brightness temperature
 power spectrum in real space is then
\begin{equation}
P_{\delta T}(k) = F \left[
P^{1b}_{\delta_{\rm HI} }(k) + P^{2b}_{\delta_{\rm HI}}(k) \right]\,,
\label{eqn:p3d}
\end{equation}
and is parameterized by the bubble radius $R$ and mean ionization $x_e$ at a given redshift
(see \fig{pk}).  
 
Like the model of \cite{Santos04}, our model is analytic and allows a rapid exploration 
of parameter space.  
Furthermore like the model of \cite{FZH04a}, 
it has the right qualitative behavior as a function 
of $x_e$ and on scales smaller than the bubble radius.
 
\begin{figure}[th]
\centerline{\epsfxsize=8.5cm\epsffile{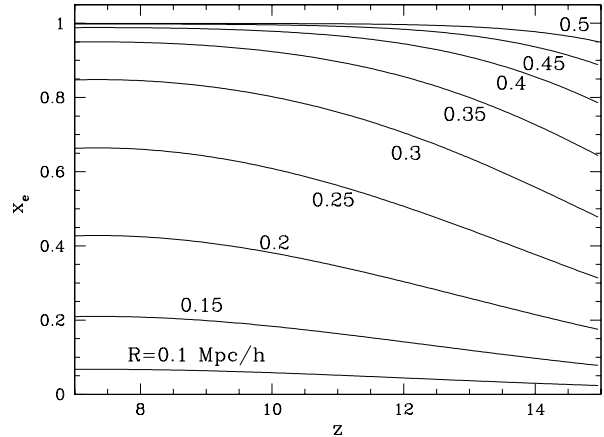}}
\caption{\label{xezvir}\footnotesize
 The condition that bubbles are seeded in halos of $T_{\rm vir} > 10^4$K sets a bubble 
 number density and hence a bubble radius $R$ given the mean ionization $x_e$.}
\end{figure}

 Despite the flexibility of our model,
 not all choices of $x_e(z)$ and $R(z)$ lead to physically plausible
reionization scenarios.   Reionization can only proceed if the seed halo
is sufficiently large for cooling and fragmentation to occur.   
  For atomic line cooling, the virial temperature of the
halo
\begin{eqnarray}
{ T_{\rm vir} \over 10^4 {\rm K}} = 1.1 \left( {\Omega_m h^2 \over 0.15} \right)^{1/3} \left( {{1+z} \over 10}\right)
\left( { M \over 10^8  M_\odot} \right)^{2/3}
\label{eqn:virial}
\end{eqnarray}
must be greater than $\sim 10^{4}$K; for molecular hydrogen cooling, $M\simgt 10^{6} h^{-1} M_\odot$
but the reionization efficiency is lower
\citep{BarHaiOst01,YosAbeHerSug03,SomBulLiv03}.    Since we are mainly interested
in redshifts $z \sim 10$ it is reasonable to choose the typical bubble radius $R$ to
correspond that required by $T_{\rm vir}(M_{\rm th}) = 10^{4}$K and $x_e$.  This is
displayed in Fig.~\ref{xezvir}.  The fiducial model is then described by a single function $x_e(z)$.

\bigskip

\section{Redshift Space Power Spectrum}
\label{pssec}

\subsection{Redshift Space Distortions}

Redshift space distortions due to the peculiar velocity of the neutral hydrogen
change the mapping between observation frequency and radial distance.
For the linear velocity field, convergences in the velocity field are associated with over dense
regions and hence enhance the apparent clustering of the density field \citep{Kaiser87}.  
The only difference between the neutral hydrogen field and the familiar density field example
is that
the neutral hydrogen can be an anti-biased tracer of the density field in the two bubble
regime 
(see \sec{sec:twobubble}).  
In this case, redshift space distortions suppress
rather than enhance the power.   In general the redshift space power spectrum $P^{s}$ of
a biased tracer $X$ of the linear power spectrum becomes
\begin{equation}
P^{s}_{X}(\vec{k}) = L^{2}(k_{\parallel}/k,b_{X}) b_{X}^{2} P_{\delta}(k) \,,
\end{equation}
where $k_{\parallel}$ is the line of sight component of $\vec{k}$ and
\begin{equation}
L(\mu,b) \equiv  1 + {d \ln D_{G} \over d \ln a} {\mu^{2} \over b} \,.
\end{equation}
Here $D_{G}$ is the linear growth rate of the density field.  For the redshifts we 
consider, $D_{G}\propto a$ and $d\ln D_{G}/\ln a= 1$.

 \begin{figure}[h]
\centerline{\epsfxsize=8.5cm\epsffile{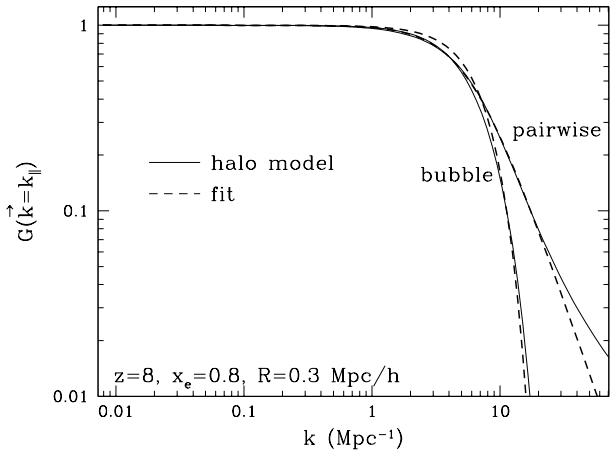}}
\caption{\label{fit}\footnotesize%
Comparison of the Fingers of God fitting functions $G_p$ (pairwise; exponential) and $G_b$ (bubble;
Gaussian)
and the halo model prediction.  The functions are normalized at $G=0.1$ such that at 
$z=8$, $\sigma_p= 29$km/s and for bubbles with $R=0.3$ Mpc/h and $x_e=0.8$, $\sigma_b
=22$km/s.}
\bigskip
\end{figure}

On small scales, peculiar velocities are associated with the virialized  motion of the
gas leading to a suppression of radial power known as the Fingers of God (FoG).
We adopt a halo model for the FoG suppression \citep{She96,Whi01}.
The 1D velocity dispersion of a halo of mass $M$
\begin{equation}
\sigma^2_h(M) = G \left( {\pi \over 6} M^2 \rho_{\rm m} \Delta_{\rm vir}\right)^{1/3} \,,
\end{equation}
where $\Delta_{\rm vir} = 18\pi^{2}$ is the virial over-density at high redshift. 
This velocity dispersion translates into smoothing across a radial spatial scale of
\begin{eqnarray}
\Delta D_{\sigma} &\approx& {{\delta z} \over H(z)} = {1 \over a H(z) } \sigma_h\nonumber \\
&=& 0.245 {\rm Mpc} \left( {\Omega_m h^2 \over 0.15}  {{1+z} \over 10 }\right)^{-1/2}  
{\sigma_h \over 30 {\rm km/s}} \,. 
\label{eqn:radialscale}
\end{eqnarray}
In Fourier space, the convolution becomes a multiplication by a suppression factor
of 
\begin{equation}
W_\sigma(k_\parallel) = \exp\left[ - {{(k_\parallel / a H})^2 \sigma_h^2(M) \over 2} \right]\,.
\end{equation}
Halos of a range of masses contribute to the density and ionization power spectra and
so the total suppression depends on a weighted average of these factors.

As indicated above, we use a constant velocity dispersion for each halo because 
the velocity dispersion profile has little impact on redshift space distortions.  
The difference between modeling  velocities through an isothermal assumption 
or through the Jeans equation in the true dark matter profile are small.
Outside the halo scale, all that matters is the
overall mean velocity dispersion of the halo. Inside the halo scale, there is
only a minor difference, the effect of taking a realistic $\sigma_h(r)$ 
 is a rescaling of a constant $\sigma_h$ within $\pm 10$\% \citep{Tinker05} .

For the density fluctuations, we separate the one and two halo contributions
as usual
\begin{equation}
P_{\delta}^s(\vec{k}) = P^{{\rm 2h} \, s}_{\delta} + P^{{\rm 1 h}\, s}_{\delta}\,,
\end{equation}
where 
\begin{eqnarray}
	\label{eqn:redshiftdeltadelta}
P^{{\rm 1h}\, s}_{\delta}(\vec{k}) &=& P^{\rm lin}_{\delta}(k)
	         \Big[ \int {d M \over M} \left( { M \over \rhom } \right) 
		{dn_h \over d\ln M}
		b(M) \nonumber\\
		&& \times W_{\sigma}(k_\parallel) y(k;M)\Big]^2 \,, \label{eqn:Psdelta}\\
P^{{\rm 2h}\, s}_{\delta}(\vec{k})&=& \int {d M \over M} \left( { M \over \rhom } \right)^2
	{d n_h\over d\ln M} y^2(k;M) W_{\sigma}^2(k_\parallel)\,, \nonumber
\end{eqnarray}
where $y$ is the Fourier transform of the gas density profile normalized such that
$y(0)=1$.  We take the gas density to trace the dark matter density in the translinear
regime considered here. The dark matter density profile can be described by
\citep{NavFreWhi04}
\begin{equation}
\rho(r) \propto {1 \over rc/r_{\rm vir}(1+ r c/r_{\rm vir})^2}\,,
\end{equation}
where the virial radius encloses the halo mass $M$ at the virial over-density
$\Delta_{\rm vir}$ and the concentration
 \citep{Buletal01}
\begin{equation}
c(M) = {9 \over {1+z} } \left( {M \over M_*} \right)^{-0.13}\,,
\label{eqn:concentration}
\end{equation}
where $M_{*}$ is defined by  $\sigma_{\rm lin}(M_*;z=0)=1.698$.

Let us define the suppression factor as 
\begin{eqnarray}
G_p(\vec{k}) = { P_{\delta}^s(\vec{k}) \over P_{\delta}(k)} \,,
\end{eqnarray}
where the real space power spectrum follows 
\eq{eqn:redshiftdeltadelta} 
with $W_\sigma \rightarrow 0$.  
The suppression 
can be well approximated by \citep{She96}
\begin{equation}
G_p(\vec{k}) \approx {1 \over 1 + (k_\parallel/aH)^2 \sigma_p^2} 
\end{equation}
or the functional form taken by an exponential distribution of pairwise velocities with
dispersion $\sigma_p$.  To extract $\sigma_{p}$, we match the numerical calculation to the fitting form
at $G_{p}=0.1$.  An example of the accuracy of the approximation is shown in 
\fig{fit}.

We follow the same procedure with FoG redshift space distortions of the one and two
bubble contributions to the ionization fluctuations.  The only difference is that in the integrals
over masses in 
\eq{eqn:Psdelta} 
the lower limit is set to $M_{\rm th}$ and the density profile $y(k;M)$ is replaced by the
bubble profile $W_{R}(k)$.
Again we can define the suppression factor as the ratio of redshift to real space power spectra.  
Since most of the contributions come from bubbles near the mass threshold,
 the suppression factor is better fit to a Gaussian
\begin{equation}
G_b({\vec k}) = \exp\left[ -{1\over 2} (k_\parallel/aH)^2 \sigma_b^2 \right] \,.
\end{equation}
We find $\sigma_{b}$ by matching this to the numerical results at $G_{b}=0.1$ and
show an example  in 
\fig{fit}.

Combining the linear redshift space distortions with the ionization and density fluctuation
FoG distortions gives the final form for the redshift space power spectrum
 \begin{eqnarray}
 \label{eqn:p3ds}
P_{\delta_{\rm HI}}^s(\vec{k}) &=& (1- x_{e})^{2} \tilde  b_{\rm eff}^2
L^{2}(\mu,\tilde b_{\rm eff}) G_p(\vec{k}) P(k)\\
&& + ( x_{e} -  x_{e}^{2} ) G_p(\vec{k}) 
 \left[ L^{2}(\mu,1) \tilde P(k) + V_{b}\tilde W_{R}^{2} \right] \,, \nonumber
 \end{eqnarray}
 where the effective bias is
 \begin{eqnarray}
\tilde b_{\rm eff} &=& \ln(1-x_{e})b \tilde W_{R} + 1  
\end{eqnarray}
and we have rescaled the window function to absorb the differences in the
FoG suppression for density and ionization fluctuations
\begin{eqnarray}
\tilde W^2_R(\vec{k}) &=& {G_b({\vec k})\over G_p(\vec k)}  W_R^2(k) \,.
\end{eqnarray}
Note that the full redshift space power spectrum of the temperature fluctuations
\begin{equation}
P_{\delta T}^{s}(\vec{k})  = F P_{\delta_{\rm HI}}^s(\vec{k}) 
\label{eqn:redshiftspacetemperature}
\end{equation}
 is still defined by the cosmology
and two parameters $R,x_e$.   With a restriction to halos that can cool by atomic
lines in the fiducial model 
(see \subsec{sec:fiducial}), 
this reduces to a single parameter $x_e$ per redshift.  Conversely, our functional
form can be used to model an even wider range of physical conditions if $\{x_e,R,\sigma_p,\sigma_b\}$ 
are all taken as free parameters of the model and the bubble profile 
$W_R$ generalized.  Note that if $W_R$ is adjusted, $V_b$ should be consistently
modified.   Finally, as discussed in the Appendix,
to account for a distribution in bubble sizes one can replace $W_R$
with $\langle W_R \rangle$ in two bubble terms and $W_R^2$ with $\langle W_R^2 \rangle$  in one bubble
terms which would also eliminate oscillatory artifacts in the power spectrum.

\smallskip

\subsection{1D power spectrum}
\label{rsigmasec}

Foreground contamination is perhaps the largest problem facing all 21cm experiments.
Since foregrounds tend to be smooth in frequency, it has long been realized that one of the
most robust observables will be the radial structure in the temperature field \citep{HogRee79,ScoRee90,GnedinShaver03,Morales05,OhMack03,xiaomin04}.  We will therefore focus 
 on the 1D power spectrum for the rest of the paper. This will also enable us to
understand intuitively the impact of redshift space distortions on future experiments.  
The modeling of the 3D power spectrum in the previous sections, however, can also be applied
to interpret foreground-cleaned data where the transverse structures can be extracted.

\begin{figure}[th]
\centerline{\epsfxsize=7.5cm\epsffile{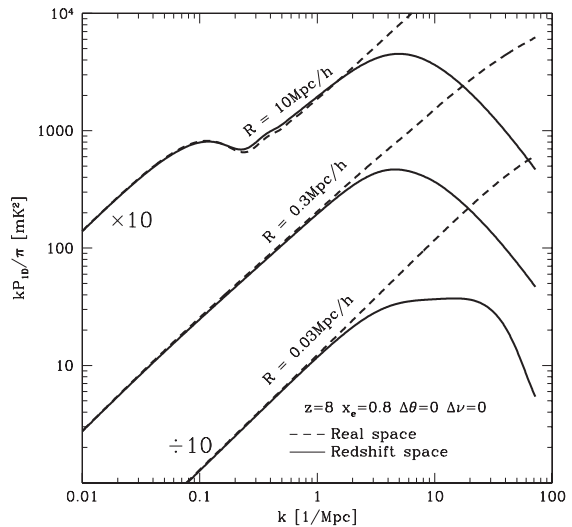}}
\caption{\label{p1d_R_Fig}\footnotesize%
Real (dashed) and redshift (solid)
space 1D power spectra for different fiducial HII patch sizes 
with perfect angular and frequency resolutions.
All parameters are the same except HII patch size $R$.  
The curves for $R=10h^{-1}\Mpc$ and $R=0.03h^{-1}\Mpc$ 
have been multiplied and divided by 10
respectively for clarity.   Note the curves represent the variance of the temperature field
in mK$^2$ contributed by a logarithmic interval in $k$.
}
\end{figure}

Employing the general relationship between 3D and 1D power spectra from \eq{eqn:1Ddef}, we
show several examples of the real and redshift space 1D power spectrum in the 21cm temperature
fluctuations  in \Fig{p1d_R_Fig}. 
We plot the variance of the temperature
field contributed per logarithmic interval in $k$ or frequency
\begin{equation}
{d \sigma_{\delta T}^2 \over d\ln k }= {k \over \pi}  P^{{\rm 1D}}_{\delta T}(k)\,,
\end{equation}
which then can be simply interpreted as the square of the amplitude of the signal. 
All curves have the same parameters except the HII bubble size $R$. Note that by fixing $\xe$ and
varying $R$, we are relaxing the constraint that bubbles are seeded by atomic cooling halos in all 
but the $R=0.3$ $h^{-1}\Mpc$  curves
(see  \Fig{xezvir}).

In real space, the bubble size $R$ places a feature in the 1D power spectrum associated
with the scale of ionization fluctuations.   For a large bubble radius $R \simgt
1$ $h^{-1}\Mpc$ 
and high mean ionization, this feature should be distinguishable from the smoother
density fluctuations even in projection.  In practice, this feature may appear as a smooth
bend in the power spectrum if the distribution of bubble sizes is wide (see Appendix).
Note that the signal
is expected to continue to rise
to smaller scales due to density fluctuations outside of the bubbles.

In redshift space, peculiar velocities
impose a second scale that cuts off the 1D power spectrum regardless of the bubble size.
 We define the
cut off scale  corresponding 
to $\sv$ and $\sb$ as $\Delta D_{\sigma}$ according to \eq{eqn:radialscale}.
For the density fluctuations
$\sigma_{p}$ controls the FoG effect, $\sv>\sb$.
All radial power on scales smaller than this cutoff scale will be erased. 

The existence of two scales in the power spectrum means that the detection of
a single feature should not automatically be associated with the bubble scale.
When $R>\Delta D_{\sigma}$, HII structures always suppress the power spectrum at
lower $k$ than the nonlinear redshift space distortion does, so that we can distinguish the HII 
bubble size 
signature from the power spectrum. With sufficient radial or frequency resolution
we will also be able to observe that FoG distortions lead to a reduction in power that
continues to increases with $k$.   

The observation of a feature followed by a cutoff in the 1D power spectrum will ensure that the feature is from HII
patches.  By measuring its evolution in redshift we obtain a statistical measure of the
evolution of the typical size of the ionized region.  However 
 in the fiducial atomic cooling model with $R=0.3$ $h^{-1}\Mpc$  only the FoG
feature is prominent.
In scenarios where $\xe$ is also adjusted, 
the prominence of the  bubble feature can be adjusted but a confusion between
the two scales can still occur.

If only
a single feature is measured and barely resolved then the ambiguity between FoG 
distortions and bubble size must be broken by measuring the transverse power spectrum.
Fortunately, the cosmological redshift space distortion or Alcock-Pacynski effect can be
considered known and does not introduce an ambiguity \citep[c.f.][]{Nusser05}. 
Foreground contamination 
in the transverse dimensions however will have to be controlled.

\section{Observed power spectrum}
\label{psdtdnsec}

\subsection{Modeling experimental resolutions}

The observed 21cm power spectrum will also be distorted by instrumental effects from finite
angular resolution $\dt$ and frequency resolution $\dn$.
These modify the 3D power spectrum as
\begin{eqnarray}
P_{\delta T}^{\rm obs}(\vec{k},\dt,\dn)  = P_{{\delta T}}^s(\vec{k}) W^2_{\dt}(k_\perp) W^2_{\dn}(k_\parallel)\,,
\label{eqn:3Dobserved}
\end{eqnarray}
and consequently the projected 1D power
spectrum through \eq{eqn:1Ddef}.
We will take the smoothing due to frequency resolution to be a Gaussian of FWHM $\dn$
so that
\begin{equation}
W^2_{\dn}(k_\parallel) = e^{- k_\parallel^2 \delta D^2 }\,,
\end{equation}
where $\delta D$ is given by 
\eq{eqn:radialdistance} 
with $\delta \nu = \dn / \sqrt{8\ln 2}$ to convert the FWHM to a Gaussian width.

For the angular resolution, a single dish experiment with a Gaussian beam of
 FWHM resolution $\dt$ would produce
a window function
\begin{equation}
W^2_{\dt}(k_\perp) = e^{- k_\perp^2 D_\perp^2 (\dt/\sqrt{8\ln 2})^2}\,,
\end{equation}
where the angular diameter distance is given by \eq{eqn:angulardiameter}.  In our fiducial
cosmology, this factor is approximately
\begin{equation}
{D_\perp \dt \over \sqrt{8\ln 2} } \approx 1.65 {\rm Mpc} 
\left[ 1 - 0.33 \left( {10 \over 1+z} \right)^{1/2} \right] \left( { \dt \over 1' } \right) \,,
\label{eqn:transversedistance}
\end{equation}
where we have also converted the angular units from radians to arcminutes on the rhs for convenience. 

All planned 21cm experiments are however interferometers.  In this case the
window function is given by a sharp cutoff in $k_\perp$ space defined by the
longest baseline.
An interferometer array with a maximum baseline of $L$ measures the transverse power spectrum
out to  
\beq{kcuteq}
k_{\rm cut} = {{2\pi L}\over {\lambda D_\pd}}\,,
\eeq 
where $\lambda$ is the observation wavelength.
With uniform coverage of the baselines, the window function can be approximated as
\begin{equation}
W^2_{\dt}(k_\perp) = \Big\{
\begin{array}{ll}
   1 &  k_\perp \le k_{\rm cut}\,, \\
   0  &  k_\perp > k_{\rm cut}\,.
\end{array}
\end{equation}
Since the net effect of this sharp cut on the 1D power spectrum 
is similar to a Gaussian beam of FWHM
\begin{eqnarray}
\Delta \theta & = & \sqrt{8\ln 2} {\lambda \over 2\pi L}\,, \nonumber \\
                       & = & \sqrt{8\ln 2} {1 \over k_{\rm cut} D_\perp} \,.
\end{eqnarray}
we will use the Gaussian window for illustrative purposes in the following sections.

\smallskip

\subsection{Observed 1D power spectra}
\label{dtdnsigmasec}

\begin{figure}[h]
\centerline{\epsfxsize=7.5cm\epsffile{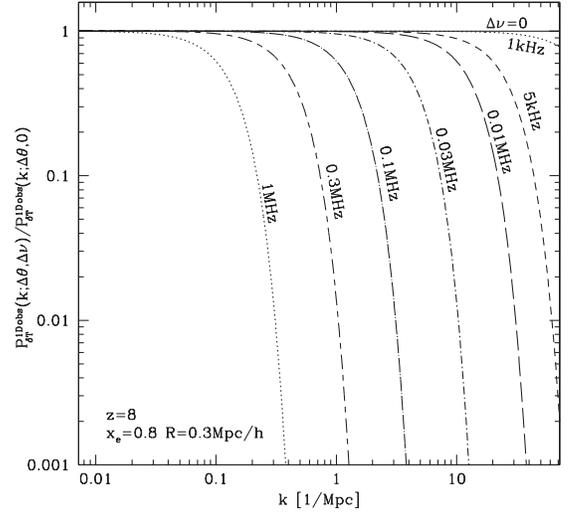}}
\caption{\label{p1dsr_diffdn_Fig}\footnotesize%
Ratio of redshift space power spectra with finite frequency resolutions $P^{\rm 1Dobs}_{\delta T}(k;\dt,\dn)$ to
the one with perfect frequency resolution $P^{\rm 1Dobs}_{\delta T}(k;\dt,0)$, for a model
with parameters $z=8$, $\xe=0.8$ and $R=0.3h^{-1}\Mpc$.
}
\end{figure}

The impact of frequency resolution on the 1D power spectrum is simple in that equations
(\ref{eqn:1Ddef}) and (\ref{eqn:3Dobserved}) imply
\begin{eqnarray}
P^{{\rm 1Dobs}}_{\delta T}(k;\dt,\dn) & =& W^2_{\dn}(k) \int {d^2 k_\perp \over (2\pi)^2}W^2_{\dt}(k_\perp)
\nonumber\\
&&\times  P_{{\delta T}}^s(\sqrt{k^2+k_\perp^2})  \,,
 \label{eqn:1Dobs}
\end{eqnarray}
so that the relative effect of frequency resolution is independent of the neutral hydrogen model.
In \fig{p1dsr_diffdn_Fig}, we plot $W^2_{\dn}$ or the ratio of 21cm redshift space 
1D power spectra for 
different channel widths $\dn$  relative to $\dn =0$.
Comparing this figure to the 
features from the ionization bubbles and the FoG distortions one might naively suppose that
the criteria for resolving these depend on $\dn$ alone.  

\begin{figure}[th] 
\centerline{\epsfxsize=7.5cm\epsffile{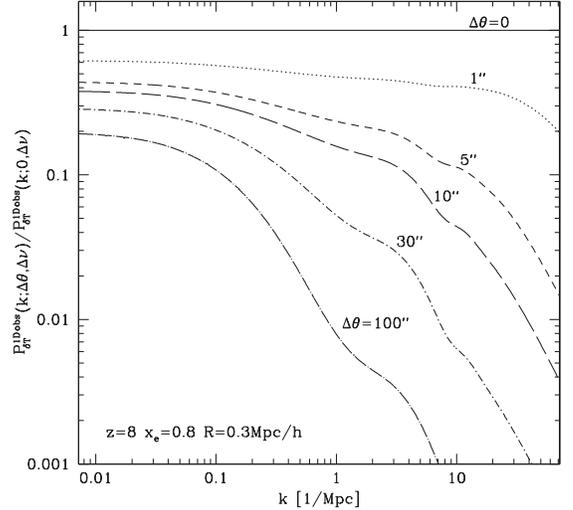}}
\caption{\label{p1dsr_diffdt_Fig}\footnotesize%
Ratio of redshift space power spectra with finite angular resolutions $P^{{\rm 1Dobs}}_{\delta T}(k;\dt,\dn)$ to
the one with perfect angular resolution $P^{\rm 1Dobs}_{\delta T}(k;0,\dn)$,   
for a model with parameters $z=8$, $\xe=0.8$ and $R=0.3h^{-1}\Mpc$. 
}
\end{figure}

However for projected 1D power spectra, the angular resolution plays an important role.  Only
modes whose three dimensional wavevector $\vec{k}$ have transverse components that
can be resolved contribute to the radial fluctuations in \eq{eqn:1Dobs}.  Thus a finite angular
resolution degrades the 1D signal on all scales and additionally can lead to a new
feature in the 1D power spectrum.  
\Fig{p1dsr_diffdt_Fig} plots the ratio of redshift
space power spectra with different finite effective beam sizes $\dt$ to the one with $\dt=0$. 
Note that the suppression effect is independent of $\dn$ but does depend on the
model 3D power spectrum.   Here we have assumed the fiducial $R=0.3$ $h^{-1}$Mpc,
$x_e=0.8$ model at $z=8$.  The feature in \Fig{p1dsr_diffdt_Fig}  near 1 Mpc$^{-1}$ is due
to the assumed bubble scale.
Even with perfect frequency resolution, the angular resolution can mask features
from the bubble scale and FoG effects.  
In the context of foreground removal, \fig{p1dsr_diffdt_Fig} implies that  for $\dt \simgt
30\arcsec$ the experimental angular resolution will place a more important limitation
on cleaning than FoG redshift space distortions.

Note that the total suppression is
simply the product of the two curves in \Fig{p1dsr_diffdn_Fig} and \ref{p1dsr_diffdt_Fig}.
The optimal frequency resolution to extract most of the power in the radial
fluctuations is then dependent on angular resolution. 
Furthermore, we can develop rules of thumb for the instrumental specifications required
to unambiguously resolve features such as the FoG effect and the bubble scale as we shall now
see.

\subsection{Resolving redshift space distortions}

For FoG redshift space distortions, the critical scale where all fluctuations will be suppressed
is associated with the pairwise velocity dispersion $\sigma_p$ since that affects the
small scale density contributions.    For the ionization fluctuations, $\sigma_b$ typically places
the suppression scale below that of the bubble scale and hence is less relevant.  
The pairwise velocity dispersion scale is independent of the bubble model
and so leads to robust criteria for its resolution.

Combining \fig{p1dsr_diffdn_Fig} with \fig{p1dsr_diffdt_Fig}, if we have both high frequency
resolution $\dn<0.03\MHz$ and high angular
resolution $\dt<30\arcsec$, which means both suppress the power at
smaller scales than the FoG effect does, we would be able to unambiguously detect the FoG effect in the
fiducial model with
instrumental sensitivity that can extract mK level signals. 
More generally,  we require that $\delta D\ll \Delta D_{\sv}$ and
$D_\perp\dt/\sqrt{8\ln2}\ll \Delta D_{\sv}$ to observe nonlinear redshift
space distortions at small scales. 

Combining \eq{eqn:radialdistance}, \eq{eqn:radialscale}, and \eq{eqn:transversedistance}, 
we obtain the rule of thumb on resolution
requirements for observing FoG effect in 21cm experiments  
\beq{dn2slimeq}
\dn \ll \dn_{\sv}=0.034\MHz\left ({{1+z}\over{10}}\right )^{-1} \left ({{\sv}\over{30\rm{km/s}}}\right ) 
\eeq
\beq{dt2slimeq}
\dt \ll \dt_{\sv}=8.9\arcsec \left[ \left( {{1+z}\over{10}}\right )^{1/2} - 0.33 \right]^{-1}
\left ({{\sv}\over{30\rm{km/s}}}\right )
\eeq
neglecting the small cosmological dependence and assuming $z \gg 1$.
Note that redshift space distortions can only be resolved when both these criteria are
satisfied.   Likewise for foreground removal, the FoG suppression becomes the limiting factor only
if both exceed this criteria.

The characteristic scales defined above decrease with redshift. 
At $z=8$ where we do all our calculations, assuming $\sv=30\rm{km}/s$, \eq{dn2slimeq} and
\eq{dt2slimeq} suggest the characteristic resolutions are 
$\dn_{\sv}=0.04\MHz$ and $\dt_{\sv}=14\arcsec$. 
These results are
consistent with \fig{p1dsr_diffdn_Fig} and \fig{p1dsr_diffdt_Fig}. 

\smallskip

\subsection{Resolving HII regions}
\label{dtdnrsec}

We can also establish rules-of-thumb for the resolution of the HII regions or 
in our model the bubble size $R$. These supplement the considerations in 
\subsec{rsigmasec} for separability from the FoG scale.   We will consequently
consider the $R=10h^{-1}\Mpc$  case where the two scales are themselves
well separated.

\begin{figure}[h]
\centerline{\epsfxsize=8.5cm\epsffile{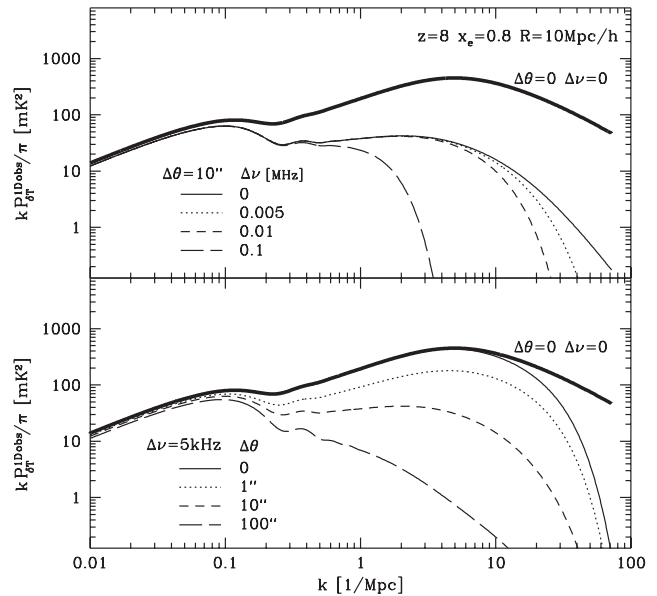}}
\caption{\label{p1d_Rx10_Fig}\footnotesize%
Redshift space 21cm 1D power spectra for different beam sizes and channel widths, assuming fiducial
HII region size $R=10h^{-1}\Mpc$.  
For comparison, we also plot  power spectra with perfect angular
and frequency resolution (thick solid lines). 
}
\end{figure}

\Fig{p1d_Rx10_Fig} shows   
how the redshift space power spectrum changes with different experimental resolutions. 
With a perfect instrument ($\dt =0$ and $\dn=0$) both the larger $R$ scale and the
FoG scale are resolved.
However, as either increases, the small scale feature is replaced by the instrumental
resolution.  
We can tell from \fig{p1d_Rx10_Fig} that for $\dn\la0.1\MHz$ and $\dt\la100\arcsec$, such large
HII structures can be
distinguished. 
It is also interesting to see that when fixing $\dt=10\arcsec$, it does not
help much to improve the frequency resolution beyond $0.01\MHz$,  
which implies the requirements for beam size and channel width
are correlated in resolving structures, as we pointed out earlier in \subsec{dtdnsigmasec}. 

%

More generally, we need both $\delta D\ll R$ and $D_\perp\dt/\sqrt{8\ln2}\ll R$
to resolve HII structures, 
\beq{dn2rlimeq}
\dn \ll \dn_{R}=0.14\MHz\left (
{{1+z}\over{10}}\right )^{-1/2} 
\left ({{R}\over{\Mpc}}\right ) 
\eeq
\beq{dt2rlimeq}
\dt \ll \dt_{R}=36.4\arcsec \left [1-0.33\left ({{10}\over{1+z}} \right )^{1/2}\right ]^{-1}\left ({{R}\over{\Mpc}}\right )
\eeq
 Either resolution not satisfying the above conditions will result in a smearing scale larger than HII
scale so that the feature will be obscured. 

Furthermore, because of the existence of redshift space distortions, $R$ and $D_\sigma$
can lead to degenerate effects
as we have already discussed in
\subsec{rsigmasec}. 
Thus the required resolutions for observing any structures, either HII
patches, or finger-of-God features, are the results of minimizing \eq{dn2slimeq} and \eq{dn2rlimeq},
plus \eq{dt2slimeq} and \eq{dt2rlimeq}, respectively. 


\smallskip

\section{Considerations for experiments}
\label{expsec}

\subsection{An ultimate experiment}
\label{idealexperimentsec}

In order to fully take advantage of all data, and go
beyond the foreground dominated region to remove foreground efficiently \citep{xiaomin04}, 
we would of course like to get as good angular and frequency resolution as possible.
On the other hand, with a reasonable ionization model like that established here there are 
guides to when an instrument will recover essentially all of the useful cosmological signal.

In the absence of redshift space distortions, we have seen that the predicted temperature
fluctuations continue to rise in rms as the frequency resolution increases due to density
fluctuations outside of the ionization bubbles.  This would imply that to be most immune to
smooth foregrounds in frequency space, one should concentrate the instrumental sensitivity in 
as small a frequency band and field as possible.  However, we have seen that FoG smoothing in
the radial direction will generically introduce a cutoff in the observed power and set
a lower limit of $\Delta \nu \sim 10$kHz on the required frequency resolution.   To match
this frequency resolution one would desire an angular resolution of $\dt = 10''$ although
there are gains all the way out to $\dt =1''$.

\begin{table}[h]
\caption{
Resolution requirements for future ``ideal" experiments \label{idealexptable}}
\begin{center}
\begin{tabular}{|c|c|c|}
\hline
 Resolve		    &$R>\Delta D_{\sv}$			                       &$R<\Delta D_{\sv}$\\
\hline
\hline
 $R$        & $\Delta\theta_{\sv}<\Delta\theta<\Delta\theta_{R}$                & Not applicable\\
                     & $\Delta\nu_{\sv}<\Delta\nu<\Delta\nu_{R}$	        & \\
\hline
 FoG       &$\Delta\theta<\Delta\theta_{\sv}(<\Delta\theta_{R})$                                 &$\Delta\theta<\Delta\theta_{\sv}$\\
       		    &$\Delta\nu<\Delta\nu_{\sv}(<\Delta\nu_{R})$                &$\Delta\nu<\Delta\nu_{\sv}$\\
\hline
Both      &$\Delta\theta<\Delta\theta_{\sv}(<\Delta\theta_{R})$                                    &Not applicable\\
 		    &$\Delta\nu<\Delta\nu_{\sv}(<\Delta\nu_{R})$                &\\
\hline
\end{tabular}
\end{center}     
\end{table}

The requirements on resolving ionization bubbles are more model dependent.
The best case scenario is that the
HII patches are larger than the characteristic scale introduced by the velocity
dispersion as one would expect near the end of reionization. 
Under this scenario,  HII patches can be resolved without ambiguities from
redshift space distortions and the experimental requirements can be
relaxed to satisfy 
$\Delta\theta<\Delta\theta_{R}$ 
and $\Delta\nu<\Delta\nu_{R}$.  
In the unfortunate scenario when $R<\Delta D_{\sv}$, \eg, if reionization ends with
the creation of many small ionization regions or the observation redshift is too high, 
radial resolution of HII structures will be difficult even with
perfect experimental resolutions.

Table~\ref{idealexptable} summarizes our angular and frequency resolutions 
considerations.  
Note that the signal to noise will also have to be high enough to detect the $\sim 10$
mK signals in question.

\subsection{First generation experiments}
\label{experimentsec}

The first generation 21 cm experiments will be far from the ultimate
angular resolution at the required signal to
noise levels considered in the previous section.  
We discuss here what the impact of redshift space modeling will be for
these experiments.

The upcoming MWA (Mileura Widefield Array)
experiment\footnote{http://web.haystack.mit.edu/MWA/MWA.html}  
has $8\kHz$ frequency resolution which is in fact sufficient for even our ultimate experiment.
At $200\MHz$ ($z=6.1$), its beam size is $3.4\arcmin$. 
The characteristic resolutions at this redshift 
for resolving HII structures and observing FoG effects are given 
by  equations (\ref{dn2slimeq})-(\ref{dt2rlimeq}) as 
$\Delta\nu_{R}=0.17\MHz (R/\Mpc)$,  
$\Delta\theta_{R}=60\arcsec (R/\Mpc)$, 
  $\Delta\nu_{\sv}=0.05\MHz (\sv/30\rm{\km s^{-1}})$, and 
 $\Delta\theta_{\sv}=17\arcsec (\sv/30\rm{\km s^{-1}})$. 
These results suggest that MWA will be able to  
observe HII patches larger than $\sim 3\Mpc$. 
It also implies that the FoG redshift space distortions will not be the limiting
factor in foreground removal as long as  the average
velocity dispersion at that time is 
less than $\sim 350\rm{km/s}$, as expected in any reasonable model. 

At $100\MHz$ ($z=13.2$), MWA can achieve an angular resolution $\sim 6.8\arcmin$.
The characteristic resolutions at this redshift are $\Delta\nu_{R}=0.12\MHz (R/\Mpc)$, 
$\Delta\theta_{R}=50\arcsec (R/\Mpc)$, 
  $\Delta\nu_{\sv}=0.02\MHz (\sv/30\rm{\km s^{-1}})$, and $\Delta\theta_{\sv}=10\arcsec
  (\sv/30\rm{\km s^{-1}})$. 
Similarly, 
it will not be able to resolve HII regions if they
are smaller than $\sim 8\Mpc$ and FoG distortions
are even less of a limiting factor for foreground removal.

Another upcoming experiment LOFAR\footnote{http://www.lofar.org}, 
 has frequency resolution
$4\kHz$, angular resolution $3.1\arcmin$ for virtual core setup at $200\MHz$, 
and $5.2\arcmin$ at $120\MHz$. 
At 200 MHz ($z=6.1$), 
it will be able to observe HII regions larger than $\sim 3\Mpc$, yet will not be
affected by FoG distortions as long as $\sv<300\km/s$.  
Similarly, at 120MHz  ($z=10.8$), 
it will resolve HII regions which are larger than $\sim 6\Mpc$.

Improving the angular resolution with longer baselines can only improve these
numbers before the  characteristic scale for FoG distortions is reached 
$\Delta D_{\sigma}=0.3\Mpc(\sigma/30\rm{km s^{-1}})$. 
 If HII bubbles are smaller than this scale at the corresponding redshift, we
are unable to unambiguously resolve them from their radial structure
no matter how good the resolutions are. This complication may impact the next generation experiment
beyond MWA and LOFAR.

In conclusion, for both LOFAR and MWA, 
the complication arising from nonlinear redshift space distortions in foreground cleaning should be
minimal. 
Both experiments  
can detect large HII structures and their correlations 
at the end of reionization epoch. Yet it could be
difficult for them to distinguish smaller patches expected at earlier times.

One should also  bear in mind that the actual ability
of individual experiments to distinguish HII regions also depends on many other
factors such as quality of foreground cleaning, treatment of systematics, signal-to-noise
ratio, \etc\ 
Experimental resolution is only one of many factors that determine how much information
from the reionization epoch can be extracted from the observational data.

\bigskip

\section{Summary}
\label{discussionsec}
In this paper, we construct a self-consistent analytic model for 21cm brightness temperature
power spectrum in the observed redshift space domain.  
In its simplest form, our model has two input parameters at each redshift,
the average ionization fraction and HII bubble size. It can be readily generalized to 
consider a distribution of bubble sizes with various profiles while maintaining the
proper scalings with ionization fraction. 
Though still a toy model of reionization, it is easy to apply to rapidly explore alternatives
and has the flexibility to accommodate various physical reionization mechanisms, 
including those based on atomic line cooling.  

Utilizing this model, we study the 1D power spectrum along the line of sight.
Given that foreground contamination is expected to be smooth in frequency space,
the 1D structure of the brightness temperature is the most robust observable and
a likely stepping stone for constructing the foreground cleaned 3D maps.

We show the existence of redshift space distortions will eventually limit our ability to observe epoch of reionization.
When the average size of HII regions is smaller than the characteristic scale of the average velocity
dispersion, as is likely toward the beginning of reionization, they can not be radially resolved.
Furthermore at the low ionization levels typical of the beginning of reionization, the brightness
fluctuations will be dominated by density fluctuations.  Features due to the presence of
HII regions may be masked by even relatively small uncertainties in the redshift 
space distortions.

Combining the radial structure induced by
the HII bubbles and nonlinear redshift space
distortions, we outline criteria for planning the angular and frequency resolution
of experiments.
Even in the 1D spectra, angular resolution enters by eliminating contributions from
modes that are not purely radial thus degrading the signal.  It is a serious limiting
factor for the first generation experiments such as MWA and LOFAR.  For such 
experiments, only HII bubbles that are larger than a few Mpc can be radially
resolved even with ideal frequency resolution.  On the other hand, nonlinear
redshift distortions will not seriously affect such experiments.  They do however suggest that
for future experiments a frequency resolution of $\sim 10$kHz and an angular resolution
of 10\arcsec\ will be sufficient to extract most of the information from the radial power spectrum
near the end of reionization.

\bigskip

\acknowledgments

The authors wish to thank Nick Gnedin, Andrey Kravtsov, 
Miguel F. Morales, Michael Mortonson, Mario Santos, Yong-Seon Song, and Jeremy Tinker 
for helpful discussions and comments.
XW is supported by 
the Kavli Institute for Cosmological Physics through the grant NSF PHY-0114422.   
WH is supported by the DOE under contract DE-FG02-90ER-40560
and the Packard Foundation.
\bigskip

\appendix

\section{Distribution of Bubble Sizes}
\label{Rdist}

A distribution of bubble sizes can be straightforwardly accommodated in the framework
of our model following \cite{Mortonson06}.  Let us denote the probability of obtaining
a bubble of radius within $dR$ of $R$ as $P(R)dR$.  The ionization fraction associated with
this bubble distribution is given by  
\begin{eqnarray}
\label{xevavg}
x_{e} &=& 1 - e^{-\bar n_{b} \langle V_{b} \rangle} \,,
\end{eqnarray}
where the average bubble volume is 
\begin{eqnarray}
\expec{V_b}&=&\int dRP(R)V_b(R) \,.
\end{eqnarray}
The window functions in one and two bubble term power spectra calculations 
are also replaced by average values. 
For two bubble terms, $W_R$ in \eq{eqn:p2b} becomes 
\beq{wravg}
\expec{W_R}={1\over{\expec{V_b}}}\int dRP(R)V_b(R)W_R \ .
\eeq
For one bubble terms, $W_R^2$ in \eq{eqn:onebubble} is replaced by
\beq{vbwr2avg}
\expec{W_R^2}={1\over{\expec{V_b}^{2}}}\int dRP(R)V^2_b(R)W^2_R \ .
\eeq
Note that the approximation to the convolution term in \eq{eqn:convolvelimit}
 still holds with $\sigma_{R}^{2}$ becoming
\begin{equation}
\langle \sigma_{R}^{2} \rangle = {1\over{\expec{V_b}^{2}}}\int dRP(R)V^2_b(R)\sigma_{R}^{2} \,.
\end{equation}

There are two qualitative effects of including a distribution.  The averaging process
removes the ringing of the 3D power spectra and broadens the feature associated
with the bubble size.  It also can change the relative strengths of the one and two bubble
contributions.  For example, since
\begin{equation}
\lim_{k\rightarrow 0}
{\expec{W_{R}^{2}} \over \expec{W_{R}}^{2}}
 \rightarrow 
 {\expec{V_{b}^{2}} \over \expec{V_{b}}^{2}} \,,
\end{equation}
the larger bubbles in the distribution enhance the power in the one bubble
term at low $k$.

To make these considerations concrete, let us consider a lognormal distribution 
\beq{rdisteq}
P(R)={1\over R}{1\over{\sqrt{2\pi}\sigma_{\ln R}}} 
\exp{ \left [{{(\ln R-\ln R_0 + 3\sigma^{2}_{\ln R}/2)^2}\over{2\sigma^2_{\ln R}}} \right ] }\,,
\eeq
where the offset is defined so that $\langle V_{b}\rangle = V_{0} = 4\pi R_{0}^{3}/3$
and similarly, $\expec{V_b^2}=V_0^2\exp{(9 \sigma^2_{\ln R})}$.
Numerical simulations suggest that  a reasonable range for  $\sigma_{\ln R}$ is $[0.5,1]$
\citep{Furlanetto05}.

\begin{figure}[th] 
\centerline{\epsfxsize=8.cm\epsffile{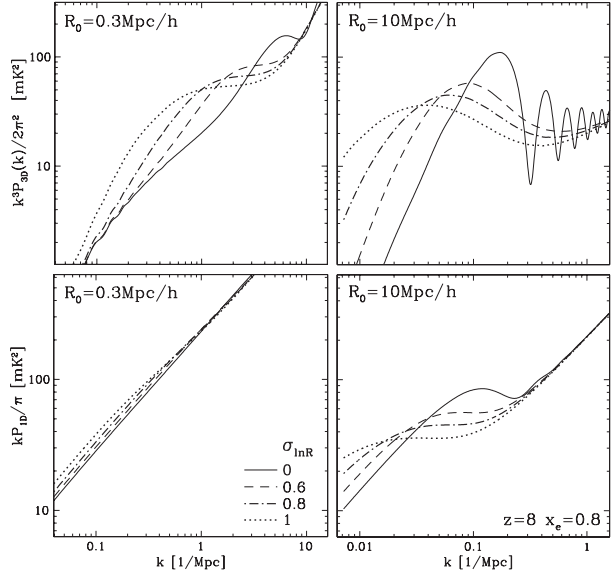}}
\caption{\label{rdist_Fig}\footnotesize%
Dependence of 3D (top) and 1D (bottom) 
real space power spectra on the width of the bubble radii distribution 
$\sigma_{\ln R}$ for two choices of $R_0 = 0.3$Mpc$/h$ (left) and $10$Mpc$/h$ (right).
Note that oscillations in the 3D spectra are averaged out and the feature due to the bubbles
is drawn out and extends to larger scales.  Models are evaluated at $z=8$ with $\xe=0.8$.
}
\end{figure}

In \fig{rdist_Fig}, we show the 1D and 3D real space power spectra for several
values of $\sigma_{\ln R}$ and two choices of $R_0$.  
In the 3D case, all oscillations are averaged away and also the bubble transition takes
place over a wide range of scales extending out to  
$\ln R \approx \ln R_0 + 3\sigma_{\ln R}^2/2$ due to the scaling of $\expec{V_b^2}$.  
Analogous effects occur for the 1D power spectra and the shift in the transition scale toward
larger scales
can make the bubble transition slightly more prominent.
Nevertheless the qualitative conclusions of the main paper remain unchanged once an allowance has been
made for the effective characteristic bubble scale.

\bigskip


\ed